# New Attack Strategy for the Shrinking Generator

Pino Caballero-Gil[1], Amparo Fúster-Sabater[2] and M. Eugenia Pazo-Robles[3]

[1] Faculty of Mathematics. University of La Laguna, 38271 La Laguna, Tenerife, Spain.
pcaballe@ull.es
[2] Institute of Applied Physics. Spanish High Council for Scientific Research. Serrano, 144, 28006, Madrid, Spain
amparo@iec.csic.es
[3] Argentine Business University, Lima 717, Buenos Aires, Argentina.
mpazorobles@uade.edu.ar

**Abstract**. This work shows that the cryptanalysis of the shrinking generator requires fewer intercepted bits than what indicated by the linear complexity. Indeed, whereas the linear complexity of shrunken sequences is between $A \cdot 2^{(S-2)}$ and $A \cdot 2^{(S-1)}$, we claim that the initial states of both component registers are easily computed with less than $A \cdot S$ shrunken bits. Such a result is proven thanks to the definition of shrunken sequences as interleaved sequences. Consequently, it is conjectured that this statement can be extended to all interleaved sequences. Furthermore, this paper confirms that certain bits of the interleaved sequences have a greater strategic importance than others, which may be considered as a proof of weakness of interleaved generators.

## 1    Introduction

Stream ciphers are considered nowadays the fastest encryption procedures. Consequently, they are implemented in many practical applications e.g. the algorithms A5 in GSM communications (GSM), the encryption system E0 in Bluetooth specifications (Bluetooth) or the algorithm RC4 (Rivest, 1998) used in SSL, WEP and Microsoft Word and Excel.

From a short secret key (known only by the two interested parties) and a public algorithm (the sequence generator), a stream cipher procedure is based on the generation of a long sequence of seemingly random bits. Such a sequence is called the keystream sequence.

For the encryption, the sender computes the bitwise exclusive OR (XOR) operation among the bits of the original message or plaintext and the keystream sequence. The result is the ciphertext to be sent. For the decryption, the receiver generates the same keystream, computes the same bitwise XOR operation between the received ciphertext and the keystream sequence and obtains again the original message.

Most keystream generators are based on Linear Feedback Shift Registers (LFSRs) (Golomb, 1982), which are linear structures characterized by their length (the number of memory cells), their characteristic polynomial (the feedback function) and their initial states (the seed or key of the cryptosystem). If the characteristic polynomial is a primitive polynomial (Lidl and Niederreiter, 1986), then the LFSRs generate Pseudo-Noise sequences (*PN*-sequences) with good pseudorandomness properties.

For a survey on primitive LFSRs, *PN*-sequences and shift equivalences, the interested reader is referred to (Golomb, 1982). In stream cipher procedures, the *PN*-sequences are combined by means of nonlinear functions in order to produce keystream sequences for cryptographic application. Combinational generators, nonlinear filters, clock-controlled generators and irregularly decimated generators are just some of the most popular nonlinear sequence generators. All of them produce keystreams with high linear complexity, long period and good statistical properties (Caballero-Gil and Fúster-Sabater, 2004) (Fúster-Sabater, 2004).

Most cryptanalysis on stream ciphers are performed under a known plaintext hypothesis, that is to say, it is assumed that the attacker has direct access to a portion of the keystream sequence (the intercepted sequence). From these intercepted bits, the attacker has to deduce the cryptosystem key. Once the key is known, as the sequence generator is public, the whole keystream sequence can be reconstructed. The complexity of this attack is always compared with that of key exhaustive search and if the former complexity is lesser, the cryptosystem is said to be broken.

This work focuses on a particular kind of stream ciphers based on LFSRs: the class of shrinking generators. They are made out of two LFSRs and an irregular decimation. Shrinking generators have been thoroughly analyzed in several papers such as (Simpson et al, 1998), (Kanso, 2003) and (Caballero-Gil and Fúster-Sabater, 2006). Nevertheless, we present a new and efficient cryptanalytic attack requiring much lesser amount of intercepted bits than that of the previous attacks. The basic idea of this cryptanalysis consists in defining the output sequence of a shrinking generator as an interleaved sequence (see (Gong, 1995) and (Jiang et al, 2002)). The interleaved characteristic of the shrinking generator reveals weaknesses that lead to practical attacks. In addition, we conjecture that these weaknesses can be extended to all interleaved sequence generators with application in cryptography.

The paper is organized as follows: in section 2, the description and characteristics of the shrinking generator is introduced. Interleaved configuration and related results are developed in section 3. A cryptanalytic attack against the shrinking generator that exploits the condition of interleaved sequence is presented in section 4, while the generalization of this technique to other cryptographic interleaved generators appears in section 5. Finally, conclusions in section 6 end the paper.

## 2  The Shrinking Generator

The so-called Shrinking Generator (SG) is a nonlinear keystream generator composed by two LFSRs (Coppersmith et al, 1993) so that a control register *SRS* decimates the sequence produced by the other register *SRA*. *S* and *A* denote respectively their corresponding lengths and fulfil that *(S, A)*= 1 and *S*<*A*. $P_S(x)$ and $P_A(x) \in GF(2)[x]$ denote their corresponding primitive characteristic polynomials.

The sequence $\{s_i\}$, produced by *SRS*, controls the bits of the sequence $\{a_i\}$ produced by *SRA* which are included in the output shrunken sequence $\{z_j\}$ according to the following rule: If $s_i=1$ then $z_j=a_i$, and if $s_i=0$ then $a_i$ is discarded.

As different pairs of *SRA/SRS* initial states can generate the same shrunken sequence, in the sequel we assume, without loss of generality, that the first term of the sequence $\{s_i\}$ equals 1, that is $s_0=1$. According to (Coppersmith et al, 1994), the period of the shrunken sequence is:
$$T = (2^A - 1)\, 2^{(S-1)},$$
its linear complexity, notated *LC*, satisfies the following inequality:
$$A \cdot 2^{(S-2)} < LC \leq A \cdot 2^{(S-1)},$$
and its characteristic polynomial is of the form:
$$P_{ss}(x) = (P(x))^p$$
where $P(x)$ is an $A$-degree primitive polynomial in $GF(2)[x]$ and $p$ is an integer in the interval $2^{(S-2)} < p \leq 2^{(S-1)}$. Moreover, it can be proven (Shparlinski, 2001) that the shrunken sequence has also good distributional statistics. Therefore, due to all these good characteristics, this scheme has been traditionally used as keystream sequence generator with application in secret-key cryptography.

## 3  Interleaved Configuration

The $(2^A-1)\cdot 2^{(S-1)}$ bits of a period of any shrunken sequence $\{z_j\}$ can be arranged into a $(2^A-1) \times 2^{(S-1)}$ matrix that we will call *Interleaved Configuration* and will denote by *IC*. In fact,

$$IC = \begin{pmatrix}
z_0 & z_1 & z_2 & \cdots & z_{2^{(S-1)}-1} \\
z_{2^{(S-1)}} & z_{2^{(S-1)}+1} & z_{2^{(S-1)}+2} & \cdots & z_{2 \cdot 2^{(S-1)}-1} \\
z_{2 \cdot 2^{(S-1)}} & z_{2 \cdot 2^{(S-1)}+1} & z_{2 \cdot 2^{(S-1)}+2} & \cdots & z_{3 \cdot 2^{(S-1)}-1} \\
z_{3 \cdot 2^{(S-1)}} & z_{3 \cdot 2^{(S-1)}+1} & z_{3 \cdot 2^{(S-1)}+2} & \cdots & z_{4 \cdot 2^{(S-1)}-1} \\
\vdots & \vdots & \vdots & \vdots & \vdots \\
z_{(2^A-2) \cdot 2^{(S-1)}} & z_{(2^A-2) \cdot 2^{(S-1)}+1} & z_{(2^A-2) \cdot 2^{(S-1)}+2} & \cdots & z_{(2^A-1) \cdot 2^{(S-1)}-1}
\end{pmatrix}$$

The following result allows us to identify each element of the matrix *IC* with the corresponding term of the sequence $\{a_i\}$.

**Theorem 3.1**

The *Interleaved Configuration* of the shrunken sequence $\{z_j\}$ can be written in terms of the elements of the sequence $\{a_i\}$ such as follows:

$$IC = \begin{pmatrix}
a_{o0} & a_{o1} & a_{o2} & \cdots & a_{o(2^{(S-1)}-1)} \\
a_{(2^S-1)+o0} & a_{(2^S-1)+o1} & a_{(2^S-1)+o2} & \cdots & a_{(2^S-1)+o(2^{(S-1)}-1)} \\
a_{2 \cdot (2^S-1)+o0} & a_{2 \cdot (2^S-1)+o1} & a_{2 \cdot (2^S-1)+o2} & \cdots & a_{2 \cdot (2^S-1)+o(2^{(S-1)}-1)} \\
a_{3 \cdot (2^S-1)+o0} & a_{3 \cdot (2^S-1)+o1} & a_{3 \cdot (2^S-1)+o2} & \cdots & a_{3 \cdot (2^S-1)+o(2^{(S-1)}-1)} \\
\vdots & \vdots & \vdots & \vdots & \vdots \\
a_{(2^A-2) \cdot (2^S-1)+o0} & a_{(2^A-2) \cdot (2^S-1)+o1} & a_{(2^A-2) \cdot (2^S-1)+o2} & \cdots & a_{(2^A-2) \cdot (2^S-1)+o(2^{(S-1)}-1)}
\end{pmatrix}$$

where the additive sub-indices $o_j$ $(j= 0, 1, \ldots, 2^{S-1}-1)$ depend on the bits of the sequence $\{s_i\}$ in the following way: if $s_i= 1$, then the corresponding sub-index $o_j$ equals the sub-index $i$, $o_j= i$. All the sub-indices are taken module $2^A-1$, that is to say, the period of the sequence $\{a_i\}$.

*Proof.* Since the period of the *PN*-sequence $\{s_i\}$ is $(2^S – 1)$, the number of bits with value 1 in a period is exactly $2^{S-1}$, and all the elements of any column of *IC* come from the same term $s_i= 1$ of the *PN*-sequence, the above expression for the matrix *IC* in terms of the elements of $\{a_i\}$ is obtained. □

According to the assumption $s_0= 1$, the sub-index $o0= 0$. Next, the following result analyzes the characteristics of the columns of the matrix *IC*.

**Theorem 3.2**

The sequences $\{d_j\}= \{a_{k+oj} : k= 0, (2^S - 1), 2 \cdot (2^S - 1), \cdots, (2^A - 2) \cdot (2^S - 1)\}$ $(j= 0, 1, \cdots, 2^{(S-1)}-1)$ corresponding to the columns of the matrix *IC* are shifted versions of a unique *PN*-sequence whose characteristic polynomial is given by:

$$P_D(x) = (x+\alpha^N)(x+\alpha^{2 \cdot N})(x+\alpha^{2^2 \cdot N}) \cdots (x+\alpha^{2^{(A-1)} \cdot N}),$$

where *N* is an integer defined as $N= 2^0 + 2^1 + \cdots + 2^{(S-1)}$ and $\alpha \in GF(2^A)$ is a root of the primitive polynomial $P_A(x)$.

*Proof.* Every sequence $\{d_j\}$ corresponding to the *j*-th column of *IC* is a regular decimation of the *PN*-sequence $\{a_i\}$. More precisely, such a sequence is obtained by taking one out of $(2^S-1)$ terms in $\{a_i\}$. The primality between *A* and *S* guarantees the primality between $(2^A-1)$ and $(2^S-1)$. Thus, the decimated sequence $\{d_j\}$ is also a *PN*-sequence. In addition, as every $\{d_j\}$ is obtained from $\{a_i\}$ with a decimation ratio of value $(2^S-1)$, then its characteristic polynomial $P_D(x)$ is the polynomial of the cyclotomic coset $(2^S-1)$ in the Galois Field $GF(2^A)$ generated by the roots of the polynomial $P_A(x)$ (Caballero-Gil and Fúster-Sabater, 2006). The starting point of each $\{d_j\}$ is given by the corresponding sub-index $o_j$. □

## 4    Cryptanalytic Attack

The proposed cryptanalytic attack consists in the computation of the initial states of both registers *SRA* and *SRS*. In fact, from some known bits of the shrunken sequence we have to determine the first *A* bits $(a_0, a_1, \ldots, a_{A-1})$ of the sequence $\{a_i\}$ as well as the first *S* bits $(s_0, s_1, \ldots, s_{S-1})$ of the sequence $\{s_i\}$. This attack can be divided into two different steps. In the first one, the computation of the initial state of *SRA* is carried out. In the second step and based on the *SRA* initial state, we determine the corresponding initial state of the register *SRS*.

### 4.1    *SRA* Initial State

Previously to the computation of the initial state, the following result is introduced.

**Lemma 4.1**

Given *A* bits of the shrunken sequence corresponding to *A* successive elements of any column of *IC*, the remaining bits of such a column can be determined.

*Proof.* Theorem 3.2 determines the characteristic polynomial $P_D(x)$ of the *PN*-sequence corresponding to every column of *IC*. Thus, by knowing *A* successive bits of any column and its characteristic polynomial, the linear recurrence relationship allows us to compute the remaining bits of such a column. □

The computation of the *SRA* initial state is described in the next result.

**Theorem 4.2**

Given *A* bits of the shrunken sequence corresponding to *A* successive elements of the first column of *IC*, the bits of the initial state of the register *SRA* can be determined.

*Proof.* Lemma 4.1 shows that the knowledge of *A* successive elements of the first column of *IC* allows us to generate the remaining bits of such a column. On the other hand, from Theorem 3.1 we know that the *(n + 1)*-th element of the first column of *IC* corresponds to $a_{n \cdot (2^S - 1)}$, that is to say, the *(n · (2^S -1) + 1)*-th term of the sequence generated by the register *SRA*. Consequently, we first solve the following system of modular equations in the unknowns $n_i$

$$n_i \cdot (2^S - 1) \equiv i \mod(2^A - 1) \qquad (i = 0, 1, \ldots, (A-1)), \qquad (5)$$

and then we compute successively the $(n_i+1)$-th *(i= 0, 1, …, (A-1))* elements of the first column of *IC* in order to obtain $a_0, a_1, \ldots, a_{A}$-1, respectively. □

**4.2    *SRS* Initial State**

The computation of the *SRS* initial state is described in the next result.

**Theorem 4.3**

Given *A·S* bits of the shrunken sequence corresponding to the top-left corner *(A × S)* sub-matrix of *IC*, the bits of the initial state of the register *SRS* can be determined.

*Proof.* Firstly, from the knowledge of the *(A × S)* sub-matrix of *IC*,

$$IC = \begin{pmatrix} a_0 & a_{o1} & a_{o2} & \cdots & a_{o(S-1)} \\ a_{(2^S-1)} & a_{(2^S-1)+o1} & a_{(2^S-1)+o2} & \cdots & a_{(2^S-1)+o(S-1)} \\ a_{2 \cdot (2^S-1)} & a_{2 \cdot (2^S-1)+o1} & a_{2 \cdot (2^S-1)+o2} & \cdots & a_{2 \cdot (2^S-1)+o(S-1)} \\ a_{3 \cdot (2^S-1)} & a_{3 \cdot (2^S-1)+o1} & a_{3 \cdot (2^S-1)+o2} & \cdots & a_{3 \cdot (2^S-1)+o(S-1)} \\ \vdots & \vdots & \vdots & \vdots & \vdots \\ a_{(2^A-2) \cdot (2^S-1)} & a_{(2^A-2) \cdot (2^S-1)+o1} & a_{(2^A-2) \cdot (2^S-1)+o2} & \cdots & a_{(A-1) \cdot (2^S-1)+o(S-1)} \end{pmatrix}$$

and according to Lemma 4.1, we can deduce the remaining bits of those *S* columns. Secondly, the relative shifts among columns may be computed from the comparison between consecutive columns. Since the sequence in every column of *IC* is exactly the same but starting at different points given by $a_{oj}$, as soon as a relative shift is found, the sub-index *oj* may be easily computed. In addition, each sub-index *oj* indicates the position of the *(j + 1)*-th 1 in the initial state of *SRS* while the intermediate bits are 0's. Thus, the above procedure can be repeated for *j= 1, 2, …* till we get *oj ≥ (S-1)*. In this way, the initial state of the register *SRS* is thoroughly determined. □

## 4.3 Illustrative Example

Let us consider a shrinking generator characterized by:

(1) *SRA* with length *A= 5*, characteristic polynomial $P_A(x) = x^5 + x^4 + x^3 + x^2 + 1$ and output sequence *{$a_i$}*.

(2) *SRS* with length *S= 4*, characteristic polynomial $P_S(x) = x^4 + x^3 + 1$ and output sequence *{$s_i$}*.

(3) The characteristic polynomial of the shrunken sequence is $P_{ss}(x) = P_D(x)^p = (x^5 + x^3 + x^2 + x + 1)^8$.

Given 20 bits of the shrunken sequence corresponding to a *(5 × 4)* top-left corner sub-matrix of *IC*

$$SUB_{IC} = \begin{pmatrix} 1 & 0 & 1 & 1 \\ 1 & 0 & 0 & 1 \\ 0 & 1 & 0 & 1 \\ 0 & 1 & 1 & 1 \\ 0 & 0 & 0 & 1 \end{pmatrix},$$

we can launch a cryptanalytic attack against the shrinking generator in order to obtain the initial states of both LFSRs. Table 1 shows the computations carried out for cryptanalyzing the above described generator. The most left column represents the indices of rows in the *IC* matrix, $n_i$= 0, 1,…, 30. Next column shows the position of the terms *($a_0$, $a_1$, …, $a_4$)* of the sequence *{$a_i$}* in the first column *{$d_o$}* of the matrix *IC*, obtained from Theorem 3.1. The following columns of the Table 1 represent the matrix *IC*: in boldface the *(5 × 4)* sub-matrix with the known bits, the remaining bits of *{$d_0$}* are the bits computed to determine the initial states of *SRA* and *SRS*, and the symbols - correspond to unknown bits of the shrunken sequence.

*Computation of the SRA initial state:* According to Theorem 4.2, we compute the positions of the *($n_i$+1)*-th elements of the first column of *IC* by solving the equation system

$$n_i \cdot 15 \equiv i \mod 31 \quad (i = 0, 1, \ldots, 4).$$

That is to say, $n_0$= 0, $n_1$= 29, $n_2$= 27, $n_3$= 25 and $n_4$= 23. Then, by means of the characteristic polynomial $P_D(x)$ we determine the values of the *($n_i$+1)*-th *(i= 0, 1, ..., 4)* elements of the first column *{$d_0$}* of *IC*. Consequently, $a_0$= 1, $a_1$= 0, $a_2$= 0, $a_3$= 1 and $a_4$= 1 (see Table 1). Therefore, the initial state of the register *SRA* (1, 0, 0, 1, 1) has been determined.

*Computation of the SRS initial state:* According to Theorem 4.3, we compute the relative shifts between consecutive columns in the matrix *IC*:

- Computation of $o_1$: We know $a_1$ at the (29+1)-th position of the first column and compute its *S-1= 4* successive bits in the column. We compare these 5 bits (0, 0, 1, 1, 0) with the first 5 bits (0, 0, 1, 1, 0) of the second column *{$d_1$}* (see Table 1) and find that there is a coincidence, thus $o_1= 1$.

- Computation of $o_2$: We know $a_2$ at the (27+1)-th position of the *{$d_0$}* and compute its 4 successive bits. We compare these 5 bits (0, 1, 0, 0, 1) with the first 5 bits (1, 0, 0, 1, 0) of the third column *{$d_2$}*. There is no coincidence, thus we analyze the following bit $a_3$. We know $a_3$ at the (25+1)-th position of *{$d_0$}* and compute its 4 successive bits. We compare these 5 bits (1, 0, 0, 1, 0) with the first 5 bits (1, 0, 0, 1, 0) of *{$d_2$}* (see Table 1). There is coincidence, thus $o_2= 3$. Since $o_2= 3 \geq S - 1$, we have determined the initial state of *SRS*. In fact, $s_0= 1$, $o_1= 1$ implies $s_1= 1$, $o_2= 3$ implies $s_2= 0$ and $s_3= 1$. Therefore, the *SRS* initial state is *($s_0$, $s_1$, $s_2$, $s_3$)= (1, 1, 0, 1)*. Note that only the knowledge of three columns of the sub-matrix has been necessary to identify the initial state of *SRS*. Indeed, this number equals the number of bits 1 in the initial state of the selector register. The maximum number of known bits corresponds to *SRS* initial state with all bits 1. In the remaining cases, fewer bits are sufficient.

Once the initial states of both register are determined, the whole shrunken sequence that is the keystream sequence can be computed.

Table 1. Matrix *IC* corresponding to the described *SG*

| $n_i$ | {$a_i$} | $d_0$ | $d_1$ | $d_2$ | $d_3$ | $d_4$ | … | $d_7$ |
|---|---|---|---|---|---|---|---|---|
| 0 | $a_0$ | **1** | **0** | **1** | **1** | - | - | - |
| 1 |  | **1** | **0** | **0** | **1** | - | - | - |
| 2 |  | **0** | **1** | **0** | **1** | - | - | - |
| 3 |  | **0** | **1** | **1** | **1** | - | - | - |
| 4 |  | **0** | **0** | **0** | **1** | - | - | - |
| 5 |  | - | - | - | - | - | - | - |
| … |  | - | - | - | - | - | - | - |
| 23 | $a_4$ | 1 | - | - | - | - | - | - |
| 24 |  | - | - | - | - | - | - | - |
| 25 | $a_3$ | 1 | - | - | - | - | - | - |
| 26 |  | 0 | - | - | - | - | - | - |
| 27 | $a_2$ | 0 | - | - | - | - | - | - |
| 28 |  | 1 | - | - | - | - | - | - |
| 29 | $a_1$ | 0 | - | - | - | - | - | - |
| 30 |  | 0 | - | - | - | - | - | - |

### 4.4 Computational Features of the Attack

The computational complexity of the proposed cryptanalytic attack must be analyzed by distinguishing two different phases: off-line and on-line, each one with its corresponding computational complexity.

*Off-line computational complexity:* Corresponding to the phase that is to be executed before intercepting sequence. It includes:

- Computation of the characteristic polynomial $P_D(x)$ by means of equation (4). This computation is necessary in order to obtain the linear recurrence relationship for the terms of the *PN*-sequence.
- Computation of the positions $n_i$ *(i= 0, 1, …, A-1)* on the first column of the matrix *IC* by means of equation (5). This computation is necessary in order to determine the bits of the initial state of *SRA*.
- Computation of different elements of the extension field $GF(2^A)$, that is to say, $\alpha^{n_i}$ *(i=0, 1, …, A-1)*, by means of the Zech log table method (Assis and Pedreira, 2000) for arithmetic over $GF(2^A)$. This computation is necessary in order to determine the *A* successive elements of each $a_i$ *(i= 0, 1, …, A-1)* on the first column of the matrix *IC*.

*On-line computational complexity:* Corresponding to the phase that is to be executed after intercepting sequence. Consequently, it can be considered the computational complexity of the proposed attack in practice. According to the previous subsections, the computational method for the computation of the *SRS* initial state consists in the comparison of series of bits coming from formulated hypothesis and from intercepted bits. The comparison is carried out by means of a few bitwise logical operations on a bounded number of bits, so the worst-case computational complexity is *O(A),* linear in the length of the *SRA*.

Finally, after comparing the proposed attack with those found in the literature we get that all of them are exponential in the lengths of the registers. In particular, the complexity of the divide-and-conquer attack proposed in (Simpsom et al, 1998) is $O(2^S)$. The probabilistic correlation attack described in (Golic and O'Connors, 1995) has a computational complexity of value $O(A^2 \cdot 2^A)$. Also the probabilistic correlation attack introduced in (Johansson, 1998) is exponential in *A*. In contrast with the previous attacks, in this work a deterministic attack has been proposed that improves the complexity of the previous cryptanalytic approaches and that requires only $A \cdot S$ intercepted shrunken bits in order to be launched.

## 5  Generalization to Interleaved Sequences

First of all, we introduce the general definition of interleaved sequence (Jiang et al, 2002).

**Definition 5.1**
Let *f(x)* be a polynomial over *GF(q)* of degree *r* and let *m* be a positive integer. For any sequence *{u$_k$}* over *GF(q)*, we write $k= i \cdot m + j$ with *(i= 0, 1,…)* and *(j= 0, ..., m-1)*. If every sub-sequence of *{u$_k$}* defined as *{u$_{i \cdot m+j}$}* is generated by *f(x),* then the sequence *{u$_k$}* is called *interleaved sequence* over *GF(q)* of size *m* associated with the polynomial *f(x)*.

Table 2 shows the interleaved sequence *{u<sub>k</sub>}* over *GF(2)* associated with the 3-degree characteristic polynomial $f(x) = x^3 + x + 1$ over *GF(2)* and size *m = 4*. Reading by rows, the interleaved sequence is $\{u_k\}$ = *{1, 1, 1, 1, 1, 0, 1, 0, 0, 0, 1, 1, 0, 1, 0, 1, 1, 0, 0, 1, 0, 1, 1, 0, 1, 1, 0, 0}* while by columns the sequence is made out of *{u<sub>j</sub>}* (*j = 0, 1, 2, 3*) four shifted versions of the *PN*-sequence generated by *f(x)*.

**Table 2.** Interleaved sequence with 4 shifted versions of the same *PN*-sequence.

| $u_0$ | $u_1$ | $u_2$ | $u_3$ |
|---|---|---|---|
| 1 | 1 | 1 | 1 |
| 1 | 0 | 1 | 0 |
| 0 | 0 | 1 | 1 |
| 0 | 1 | 0 | 1 |
| 1 | 0 | 0 | 1 |
| 0 | 1 | 1 | 0 |
| 1 | 1 | 0 | 0 |

Interleaved sequences are currently used as keystream sequences with application in cryptography. They can be generated in different ways:

(1) By a LFSR controlled by another LFSR (which may be the same one) e.g. multiplexed sequences (Jennings, 1983), clock-controlled sequences (Beth and Piper, 1985), cascaded sequences (Gollmann and Chambers, 1989), shrinking generator sequences (Coppersmith et al, 1993), etc.

(2) By one or more than one LFSR and a feed-forward nonlinear function, e.g. Gold-sequence family, Kasami (small and large set) sequence families, GMW sequences, Klapper sequences, No sequences etc. See (Gong, 1995) and the references cited therein.

In brief, a large number of well-known cryptographic sequences are included in the class of interleaved sequences. Next, the link between interleaved sequences and shrunken sequences is expressed in the following result.

**Theorem 5.2**

Shrunken sequences are interleaved sequences of size $2^{(S-1)}$.

*Proof.* Let *{z<sub>k</sub>}* be a shrunken sequence with characteristic polynomial $P(x)^p$ where *P(x)* is an *A*-degree primitive polynomial and *p* is an integer in the interval $2^{(S-2)} < p \leq 2^{(S-1)}$. According to the interleaved configuration *IC*, we may express *{z<sub>k</sub>}* in terms of *m* sequences *{z<sub>j</sub>}* where $\{z_j\} = \{z_{i \cdot m + j}\}$ with $i \geq 0$, $m = 2^{(S-1)}$ and *(j = 0, ..., m - 1)*. Since by Theorem 3.2 the sequences *{z<sub>j</sub>}* are generated by the same characteristic polynomial *P<sub>D</sub>(x)*, we get that the shrunken sequence *{z<sub>k</sub>}* is an interleaved sequence of size $2^{(S-1)}$ associated with the polynomial *P<sub>D</sub>(x)*. □

The previous theorem proves that shrunken sequences are interleaved sequences. Moreover, section 4 shows that the knowledge of a number of bits of the shrunken sequence allows us to mount a cryptanalytic attack against the shrinking generator. As many cryptographic sequence generators produce interleaved sequences, then the previous considerations take us into the following conjecture:

**Conjecture 5.3**
Given a number of known bits corresponding to a top-left corner sub-matrix of the interleaved configuration *IC* of any interleaved sequence, it is always possible to obtain the whole interleaved sequence.

The confirmation of this conjecture would prove the weakness of interleaved generators for cryptographic purposes.

# 6    Conclusions

In this work a new deterministic cryptanalytic attack against the class of shrinking generators has been proposed. The amount of necessary intercepted bits and the on-line computational complexity of such an attack are much lesser than those of other standard cryptanalysis. The basic idea of the proposed attack consists in defining the shrunken sequence as an interleaved sequence, and then using the weaknesses inherent to its interleaved configuration for launching a practical attack. A direct consequence of this result is its possible generalization to other interleaved sequence generators of cryptographic purpose. Therefore, the security of all interleaved generators should be carefully checked.

**Acknowledgements**
This research has been supported by the Spanish Ministry of Science and Innovation under Project TIN2008-02236/TSI, and developed in the frame of the project HESPERIA (www.proyecto-hesperia.org) under programme CENIT supported by Centro para el Desarrollo Tecnológico Industrial (CDTI) and the companies: Soluziona, Unión Fenosa, Tecnobit, Visual-Tools, BrainStorm, SAC and TechnoSafe.